\documentclass[prd, aps, nobibnotes, nofootinbib,
eqsecnum,
preprint,
amsmath,
amssymb]{revtex4-2}

\usepackage{graphicx}
\usepackage{dcolumn}
\usepackage{bm}
\usepackage{hyperref}
\usepackage{mathrsfs}
\usepackage{xcolor}
\usepackage{slashed}
\usepackage{subfigure}

\begin{document}

\title{Spin-flux attachment by dimensional reduction of vortices}
\author{Shantonu Mukherjee}
\email{shantonumukherjee@bose.res.in}
\affiliation{S N Bose National Centre for Basic Sciences, Block JD, Sector III, Salt Lake, Kolkata 700106, India}
\author{Amitabha Lahiri}
\email{amitabha@bose.res.in}
\affiliation{S N Bose National Centre for Basic Sciences, Block JD, Sector III, Salt Lake, Kolkata 700106, India}
	
\begin{abstract}
The description of a system of vortices in terms of dual fields provides a window to new phases of the system.
It was found recently that dualizing a 3+1-d boson-fermion system leads to a system of fermions and vortices interacting via a 2-form field through a non-local term. Here we explore some consequences of such an interaction when the degrees of freedom of the system are confined to a 2+1-d space-time. 
In particular, we show that the vortices in the 2+1-d system are attached to the fermions via their non-zero spin magnetic moment in a way similar to the phenomenon of flux attachment in Chern-Simons gauge theory coupled to matter. 
We also show that such flux attached particles exhibit fractional statistical behaviour like anyons. Thus our model provides a realization of anyons without Chern-Simons theory. 
\end{abstract}

\date{\today}

\maketitle
\newpage

\section{Introduction}
Physicists have been interested in the physics of planar systems ever since the discovery of integer Quantum Hall Effect (IQHE)~\cite{Klitzing:PRL45.494}, in which a 2d electron gas exposed in a high magnetic field exhibits non-zero Hall conductivity at integer filling fractions (magnetic flux quanta per particle). Later an even more unexpected and novel phenomenon called fractional quantum hall effect~(FQHE) was seen in similar 2d semiconducting materials where non-zero conductivity is seen at certain rational filling fraction~\cite{Tsui:PRL48.1559}. While IQHE can be explained in terms of Landau levels -- quantized energy levels of electrons in an external magnetic field, completely new ideas were needed for explaining FQHE. It turns out that the explanation of FQHE  requires that the quasiparticles are anyons -- composites of electrons and vortices of quantized flux, obeying \textit{fractional statistics} which is intermediate between bosonic and fermionic statistics~\cite{Leinaas:1977fm, Wilczek:1981du, Laughlin:1983fy, Arovas:1985yb, Jain:PRL63.199, Chen:1989xs, Feldman:2021xfn, Lee:1989fw, Rao:1992aj}.
Anyons, and the associated fractional statistics, exist only in 2+1 dimensions and are also useful in the context of $P$ and $T$-violating high-$T_c$ superconductivity\cite{Laughlin:1988fs, Wen:1989yjx, Zee:1990tg, Wilczek:1990wsc}, error corrected quantum computation etc.~\cite{Nayak:2008zza, DasSarma:2005zz, Stern:2013, Rao:2016evt} Direct observation of anyons and fractional statistics have been reported recently in FQH type systems~\cite{Bartolomei:2020Sci, Nakamura:2020Nat.Ph}. These discoveries motivate further investigation into the physics of such exotic particles and systems where they can appear.  

It was found recently that in a system of vortex strings and electrons in 3+1 dimensions, there is an effective interaction between the electrons corresponding to a linear attractive potential. This can be interpreted as electrons being joined by vortex strings~\cite{Mukherjee:2021ypk}. We can thus imagine that when restricted to 2+1 dimensions, the electrons will remain bound to the flux of the vortices.
In this paper we will consider vortex strings in a 3+1 dimensional Abelian Higgs model interacting with electrons
and reduce the system to 2+1 dimensions. We shall see that this gives rise to flux attachment -- vortices of flux attached to electrons -- which in turn show anyonic behaviour. It may seem surprising that we are starting with 3+1 relativistic quantum field theory, since in condensed matter physics a non-relativistic approach should be sufficient as the energies are much lower than the rest mass of the electron. However, an important reason for using 3+1 dimensional field theory is our method of dealing with the vortex string. We describe interactions of vortex strings in terms of a 2-form gauge potential~\cite{Kalb:1974yc} which results from using a dualization procedure. The  dualized version of the 3+1 dimensional relativistic field theory is necessarily different from the dualized version of the corresponding 2+1 dimensional theory\cite{Savit:1979ny, Peskin:1977kp, Kiometzis:1995eg, Diamantini:1995yb, Dasgupta1981}. Furthermore,  electron spin plays an important role in our description -- vortices are attached to spin, or more accurately, to the magnetic dipole moment. Spin appears naturally in the Dirac equation for the electron, so it is an additional reason to consider relativistic field theory in 3+1 dimensions. 

The vortex strings we consider are Abrikosov-Nielsen-Olesen (ANO) strings~\cite{Abrikosov:1956sx, Nielsen:1973cs} in an Abelian Higgs model of a charged scalar field  coupled to the electromagnetic gauge field. The system consists of these strings interacting with itinerant electrons via the gauge field. In the absence of strings or electrons, the Abelian Higgs model in the broken symmetry phase is dual to the theory of an antisymmetric tensor gauge field (a 2-form); in the presence of strings the 2-form couples to the world sheet of the string~\cite{Allen:1990gb, Lee:1993ty}. Dualization makes the vortex strings explicit. Adding free (i.e. unpaired or itinerant) electrons to the mix causes an additional complication, the dual theory contains non-local coupling terms. The starting point is a system described by a Lagrangian
\begin{equation}
	\mathscr{L} = -\frac{1}{4}F_{\mu\nu}F^{\mu\nu} + \frac{1}{2}(D^\mu \phi)^{\dagger}D_\mu \phi + \bar{\psi}(i\gamma^\mu \partial_\mu - m)\psi
	-V(\phi^{\dagger}\phi)
	-e A_\mu \bar{\psi}\gamma^\mu\psi\,,
\end{equation}
where the symbols have their usual meanings. This is dualized by considering the path integral, assuming there are vortex strings in the system~\cite{Mukherjee:2019vmi}.  The dual Lagrangian has the form
\begin{align}
		\mathscr{L}  =  & -\frac{1}{4}F_{\mu\nu}F^{\mu\nu}+ \bar{\psi}(i\gamma^\mu \partial_\mu - m)\psi -eA_\mu\bar{\psi}\gamma^{\mu}\psi -\frac{1}{2}e M B^{\mu\nu}\varepsilon_{\mu\nu\rho\lambda} \partial^{\rho}\frac{1}{\square}\bar{\psi}\gamma^{\lambda}\psi +\frac{1}{2} v^2\partial_\mu f \partial^\mu f \notag \\ &
		\qquad \qquad + \frac{1}{12}H^{\nu\rho\lambda}\left(\frac{1}{f^2} + \frac{M^2}{\square}\right) H_{\nu\rho\lambda} 
		-\frac{v}{2} B_{\rho\lambda}\Sigma^{\rho\lambda} - V(f^2)\,.
	\label{4d-dual}
\end{align}
Here $\Sigma_{\rho\lambda}$ is the area element of the string worldsheet, $f$ is the modulus of the Higgs field scaled by its vacuum expectation value $v$\,, $V(f^2)$ is a potential which has a minimum $V=0$ at $f=1$\,, and the charge $q$ of the complex scalar has been traded for the mass parameter $M=qv\,.$  In 3+1 dimensions the theory appears to have electrons joined by vortex strings~\cite{Mukherjee:2021ypk}. We will reduce the theory to 2+1 dimensions and find that the locations of the vortices coincide with the locations of electron spin, i.e. electrons have fluxes attached to them. We will also find that such electrons behave as anyons under exchange. Typically, anyons appear in Chern-Simons gauge theories coupled to fermions~\cite{Zhang:1988wy, Zhang:1992eu, Girvin:1987fp, Frohlich:1988qh}. Here, even though the 2+1 dimensional theory is not a Chern-Simons theory, we still find anyons. (This is similar in spirit to~\cite{Liguori:1993}, but different in details.)

Let us briefly discuss what happens in Chern-Simons theory, for the purpose of comparison with what we do in this paper. The Lagrangian of Chern-Simons theory coupled to a current~\cite{Dunne:1998qy}
\begin{equation}
	\mathscr{L} = \frac{\lambda}{2}\varepsilon^{\mu\nu\lambda}A_\mu\partial_\nu A_\lambda - A_\mu J^\mu \,,
\end{equation}
leads to the following equations of motion for the electric and magnetic fields,
 \begin{align}
		B \equiv \varepsilon_{ij}\partial_i A_j =& \rho/\lambda \,\label{CS-B}\\
		\varepsilon_{ij} E_j =&  J_i/\lambda\,. \label{CS-E}
\end{align}
Thus the magnetic field at any point is proportional to the charge density, implying that magnetic flux is bound to charges. This causes the appearance of anyons, as we will see below. The charge current is proportional to the orthogonal electric field, which is a characteristic feature of Hall effect. Both of these effects are extremely important for explaining FQHE at different filling factors. For a nonrelativistic distribution of point charges $e$ at $\vec{x}_1, \vec{x}_2, \cdots$\,, the magnetic field is 
\begin{equation}
	B(x) = \frac{e}{\lambda}\sum_i \delta^2(x-x_i)\,,
\end{equation} 
so that the magnetic field is nonvanishing only on the charges each charge is attached to a flux of $\Phi = \dfrac{e}{\lambda}$\,.  So when two such point charges are moved adiabatically around one another the wave function of any of the particle changes by a Aharonov-Bohm phase
\begin{equation}\label{eq:10}
	\exp\left(ie\oint \vec{A}\cdot d\vec{x}\right)= \exp\left(i\frac{e^2}{\lambda}\right).
\end{equation}
So if the particles are exchanged adiabatically the phase induced in the wave function of each particle is
$\Delta \theta = \frac{e^2}{2\lambda}$ which can take any arbitrary value depending on the coefficient $\lambda$ of the Chern-Simons term. These point particles coupled to flux quanta thus behave like anyons, obeying fractional statistics. In this paper, we will also find flux attachment to point charges, but through their magnetic moment rather than electric charge.

The Lagrangian of Eq.~(\ref{4d-dual}) is the starting point of our paper. In Sec.~\ref{dimred} we will reduce this to a Lagrangian of fields restricted to 2+1 dimensions. In Sec.~\ref{attach} we show that this leads to magnetic flux being attached to particles, deriving an equation similar to Eq.~(\ref{CS-B}). In Sec.~\ref{anyon} we calculate what happens when such a charge is taken adiabatically around another, and find that they obey fractional statistics. We end the paper with some additional discussion of our results and comparison with anyons in Chern-Simons theory.

\section{Dimensional reduction}\label{dimred}
We wish to dimensionally reduce this to a theory of electrons and vortices in 2+1-dimensions. In Kaluza-Klein reduction~\cite{Kaluza:1921tu, Klein:1926tv, Salam:1981xd}, one of the space directions is compactified in a circle of radius $R$\, and fields can be expanded in the form $\Phi \sim \sum \Phi_n e^{i n \varphi}$ so that the $n$-th Kaluza-Klein mode has mass given by $m_n^2 = n^2/R^2\,.$ If $R$ is sufficiently small, all modes other than the zero mode are heavy enough to be ignored in low energy calculations. Then only the zero modes of the fields propagate, and only in the uncompactified dimensions, {as the zero modes do not depend on the compact dimension}. We do not have a compactified dimension -- in our case, the reduction is done by assuming that the fields do not depend on the third space dimension~\cite{Govindarajan:1984ff}. 

Writing the 2+1-dimensional indices as $a, b, c,$ etc. and $\epsilon_{abc3} = \epsilon_{abc}\,,$ we can write for the different terms in Eq.~(\ref{4d-dual}) as
\begin{align}\label{reduction.F+psi}
	    -\frac{1}{4}F_{\mu\nu}F^{\mu\nu} &= -\frac{1}{4}F_{a b}F^{a b} + \frac{1}{2}\partial^a \phi \partial_a \phi \\
	     \bar{\psi}(i\gamma^\mu \partial_\mu - m)\psi -eA_\mu J^\mu &= \bar{\psi}(i\gamma^a \partial_a - m)\psi -eA_a J^a - e\phi J^3\,,
\end{align}
where we have written $A_3 = \phi\,$  and $J^\mu = \bar\psi\gamma^\mu\psi\,.$ For the nonlocal terms in the Lagrangian, we note that $\Box^{-1}$ is the Green function for the four dimensional wave operator, so if all functions are independent of one coordinate, it becomes the Green function for the three dimensional wave operator, which we will write as $\Delta^{-1}\,.$ Thus 
\begin{equation}\label{reduction.B-psi}
	 \frac{1}{2}e M B^{\mu\nu}\varepsilon_{\mu\nu\rho\lambda} \partial^{\rho}\frac{1}{\Box} J^\lambda = \frac{1}{2}e M B^{a b}\varepsilon_{a b c} \partial^{c}\frac{1}{\triangle}J^3 + eM B^a \varepsilon_{abc}\frac{1}{\triangle}\partial^b J^c\,,
\end{equation}
where we have now defined $B^a \equiv B^{a3}\,.$ If we also write $(\partial^a B^b - \partial^b B^a) = H^{a b}\,,$ the quadratic term for $B$ can be written as
\begin{equation}
		\frac{1}{12}H^{\nu\rho\lambda}\left(\frac{1}{f^2} + \frac{M^2}{\Box}\right) H_{\nu\rho\lambda} 
		 = -\frac{1}{4f^2}H_{a b}^2 + \frac{1}{12f^2}H_{abc}^2 - \frac{M^2}{4}H^{a b}\frac{1}{\triangle}H_{a b} + \frac{M^2}{12}H^{abc}\frac{1}{\triangle}H_{abc}
\end{equation}
The vortex string world sheet is defined as $\Sigma^{\mu\nu} = \varepsilon^{\mu\nu\rho\lambda}\partial_\rho\partial_\lambda \theta_s$ where $\theta_s$ is the singular part of the phase of the complex scalar field. We have also taken the string to be infinitesimally thin, and set $f$ to its vacuum value $f=1$ everywhere outside the string. 
If we define $\Sigma^a = (\Sigma^{03},\Sigma^{13}, \Sigma^{23}) $\,, we can write 
\begin{equation}
	\frac{v}{2}B_{\mu\nu}\Sigma^{\mu\nu}= v B_a \Sigma^a + \frac{v}{2}B_{ab}\Sigma^{ab}\,.
\end{equation}
%
%
%
%
%
The action is the result of integrating this Lagrangian over a four-dimensional volume. When we dimensionally reduce the theory, all fields are taken to be independent of the $z$-coordinate, so the integration over $z$ contributes only a length factor. We assume that the system extends in the $z$ direction by a small length $L$\,, so the fields of the three-dimensional Lagrangian are scaled by
\begin{equation}
 	A_a\rightarrow \sqrt{L}A_a\,, \phi\rightarrow \sqrt{L}\phi\,, \bar{\psi} \rightarrow \sqrt{L} \bar{\psi}\,, B_a\rightarrow \sqrt{L} B_a\,, B_{ab} \rightarrow \sqrt{L} B_{ab}\,.
\end{equation}
The charge is also scaled as
\begin{equation}
	e \to \kappa e\,,
\end{equation}
where we have written $\kappa = 1/\sqrt{L}\,.$
We note that the fields $\phi$ and $B_{ab}$ couple to $J^3$\,, which was the third component of the conserved fermionic current in three space dimensions.  Then the Lagrangian, obtained by setting all fields to be independent of the third direction, can be written as a sum of two terms, 
\begin{equation}\label{Lsplit}
	\cal{L} = \mathscr{L}_\mathrm{2+1} + \mathscr{L}_\mathrm{3}
\end{equation}
In this Lagrangian, $\mathscr{L}_\mathrm{3}$ contains all the terms involving $\phi$\,, $B_{ab}$\,, and $J^3$\,,
\begin{equation}\label{L3}
\mathscr{L}_\mathrm{3}	 = \frac{1}{2}\partial^a \phi \partial_a \phi - \kappa e\phi J^3 + \frac{1}{12}H_{abc}^2
	 + \frac{M^2}{12}H^{abc}\frac{1}{\triangle}H_{abc}+\frac{1}{2}\kappa e M B^{a b}\varepsilon_{a b c} \partial^{c}\frac{1}{\triangle}J^3 + \frac{1}{2} v B_{ab} \Sigma^{ab}\,,
\end{equation}
where we have absorbed the scaling of the fields by a redefinition, but kept $\kappa$ explicit where it multiplies 
the electronic charge $e\,.$
The remaining terms which come from the reduction are grouped together in $\mathscr{L}_\mathrm{2+1}$\,,
\begin{align}
\mathscr{L}_\mathrm{2+1} =& -\frac{1}{4}F_{a b}F^{a b}  + \bar{\psi}(i\gamma^a \partial_a - m)\psi - \kappa eA_a\bar{\psi}\gamma^{a}\psi -\frac{1}{4}H_{a b}^2   - \frac{M^2}{4}H^{a b}\frac{1}{\triangle}H_{a b}\notag \\ & \qquad + \kappa eM B^a \varepsilon_{abc}\frac{1}{\triangle}\partial^b J^c + v B^a \Sigma_a\,.
\end{align}
It is this ``reduced Lagrangian'' $\mathscr{L}_\mathrm{2+1}$ which will lead to flux attachment and fractional statistics of particles as we shall see in the following sections. The part $\mathscr{L}_\mathrm{3}$ will not contribute to upcoming discussions and hence we shall ignore that part for the remainder of this paper, coming back to talk about it briefly in Sec.~\ref{disc}.

\section{Flux attachment}\label{attach}
In the derivation of a Landau-Ginzburg type action for FQHE~\cite{Zhang:1992eu, Zhang:1988wy} a system containing nonrelativistic electrons in two spatial dimension is transformed to a bosonic system by a singular transformation of the many body fermionic wave function,
\begin{equation}
	\psi \rightarrow \tilde{\psi} = \exp{\left(-\sum_{a<b} \frac{\theta}{\pi} \alpha_{ab}\right) \psi}\,.
\end{equation}
Here $\alpha_{ab}$ is the angle made by the vector connecting particle $a$ and particle $b$ with some arbitrarily chosen reference vector, \textit{e.g.} the $X$-axis. For $\theta= (2n+1)\pi$\,, the transformed wave function $\tilde{\psi}$ is symmetric under particle exchange, and it becomes a bosonic field in second quantization. The transformed Hamiltonian contains a minimum coupling of this field with a gauge field $\vec{a}$ defined as
\begin{equation}\label{a-field}
	\vec{a}(\vec{x}_a)= \frac{\phi_0}{2\pi}\frac{\theta}{\pi}\sum_{b\neq a} \vec{\nabla} \alpha_{ab}\,,
\end{equation}
where $\phi_0= \frac{hc}{e}$ is the basic unit of quantized magnetic flux. In the second quantized formalism Eq.~(\ref{a-field}) reads 
\begin{equation}
	a_i = - \frac{\phi_0}{2\pi}\frac{\theta}{\pi} \varepsilon_{ij}
	\int d^2y \frac{x^j - y^j}{|\vec{x}-\vec{y}|^2} \rho(y)\,,
\end{equation}
where $\rho$ is the charge density. This is nothing but the solution of the differential equation
\begin{equation}
	\varepsilon^{ij} \partial_i a_j = \frac{\theta}{\pi} \phi_0 \rho\,, 
\end{equation}
which happens to be the equation of motion for  Chern-Simons theory coupled to a matter current. This equation suggests that in the bosonic phase and for $\theta= (2n+1)\pi$\,, each point particle is attached to $(2n+1)$\, flux quanta. Alternatively, one think of each vortex as carrying a net charge of $\dfrac{e}{(2n+1)}$ i.e. individual vortices have fractional charge~\cite{Paul:1986ix, Khare:1997wu}. This is called flux attachment in the context of FQHE. We will now show that a similar relationship can derived from the model we have considered, namely that of flux strings and itinerant electrons, reduced to 2+1 dimensions.

What we are looking for is the interaction between fermions and vortices (points where the flux strings cross the plane). For the 3+1 dimensional situation of Eq.~(\ref{4d-dual}) the interaction between fermions and flux strings was calculated in~\cite{Mukherjee:2019vmi} by integrating out the gauge field $B_{\mu\nu}$\,.
Therefore let us now integrate over $B_\mu$ in the reduced Lagrangian $\mathscr{L}_\mathrm{2+1}$ (we will use Greek letters for 2+1 dimensional indices). Considering only the relevant terms and adding a gauge-fixing term $-\frac{1}{2\xi}(\partial_\mu B^\mu)^2$\,, we can write this integral as
\begin{align}
		\mathcal{Z}_B & = \int \mathscr{D}B_\mu \exp\left(i\int d^3x\left(-\frac{1}{4}H_{\mu\nu}\left(1 + \frac{M^2}{\triangle} \right)H^{\mu\nu} + \kappa eM  \varepsilon^{\mu\nu\lambda}B_\mu\frac{1}{\triangle}\partial_\nu J_\lambda + v B_\mu \Sigma^\mu - \frac{1}{2\xi}\left(\partial_\mu B^\mu\right)^2\right)\right)\, \notag \\ 	
		 & =  N \exp\left(-\frac{i}{2}\int d^3x d^3y\left(J_\mu^T(x) \Delta^{\mu\nu}(x-y)J_\nu^T(y)\right)\right)\,, 
\end{align}
where we have written 
$J^\mu_T= \kappa e M\epsilon^{\mu\nu\lambda}\frac{1}{\triangle}\partial_\nu J_\lambda + v  \Sigma^\mu \, $ for the total vorticity current which couples to $B_\mu\,,$ and 
$\Delta_{\mu\nu}$ is the inverse of the quadratic operator in the action for $B_\mu\,,$
\begin{equation}
	\Delta_{\mu\nu}= \frac{1}{\triangle + M^2}\left(g_{\mu\nu} - \frac{1}{\triangle}\left(1 -\xi \left(1 + \frac{M^2}{\triangle}\right)\right)\partial_\mu\partial_\nu\right)\,.
\end{equation}

It will be convenient to split $\Sigma_\mu$ into two parts -- one that comes from non-contractible loops of the scalar field, and one that is due to any externally applied magnetic field which may be present. For the part coming from the scalar field, let us write $C_\mu \equiv \partial_\mu \theta_s$ and treat $C_\mu$ as the dynamical variable. It is the curl of $C_\mu$ which measures the flux through the vortex string -- we will relate it to the magnetic moment density of the fermions later. We will write the contribution due to the external magnetic field as $\Sigma^{{\rm e}\mu}$\,, thus the total vorticity is $\Sigma^{\mu}= \varepsilon^{\mu\nu\lambda}\partial_\nu C_\lambda + \Sigma^{{\rm e}\mu}$\,. Then we get
\begin{align}
\mathcal{Z}_B = & N \exp\left(-\frac{i}{2}\int d^3x \left({v^2}\Sigma_\mu^{\rm e}\frac{1}{\triangle+ M^2}\Sigma^{{\rm e}\mu}+ {2 \kappa ev M}\Sigma^{{\rm e}\mu}\frac{1}{\triangle +M^2}\tilde{J_\mu}+ \kappa^2 e^2 M^2\tilde{J_\mu}\frac{1}{\triangle + M^2}\tilde{J^\mu}\right.\right.
\notag \\ &
+ \left.\left. {v^2}\varepsilon^{\mu\nu\lambda}G_{\nu\lambda}\frac{1}{\triangle + M^2}\Sigma_\mu^{\rm e} 
+ 2\kappa evM C_\mu \frac{1}{\triangle + M^2} \varepsilon^{\mu\nu\lambda}\partial_\nu \tilde{J_\lambda} +\frac{v^2}{2}G_{\mu\nu}\frac{1}{\triangle + M^2}G^{\mu\nu}\right)\right)\,,
\label{Z_B}
\end{align}
where we have also defined $G_{\mu\nu}= \partial_\mu C_\nu-\partial_\nu C_\mu$\, and $\tilde{J_\lambda}= \varepsilon_{\lambda\rho\sigma}\partial^\sigma \frac{1}{\triangle}J^\rho$\,.
We now consider only the terms involving $C_\mu\,,$ at low energies. Then the propagator behaves as $\frac{1}{M^2}\,,$ so we can write those terms as
%
%
%
%
\begin{equation}
	\mathscr{L}^\prime = -\frac{1}{4}G_{\mu\nu}G^{\mu\nu} + \kappa e C_\mu  J^\mu -  \frac{1}{q}C_\mu \varepsilon^{\mu \nu\lambda}\partial_\nu \Sigma_\lambda^{\rm e}\,,
\end{equation}
after rescaling $C_\mu\rightarrow {q}C_\mu\,$ and using $\tilde{J_\lambda}= \varepsilon_{\lambda\rho\sigma}\partial^\sigma \frac{1}{\triangle}J^\rho\,.$
This Lagrangian is the same as that for an ordinary gauge field with both $J_\mu$ and $\varepsilon^{\mu \nu\lambda}\partial_\nu \Sigma_\lambda^{\rm e}$ as sources. The equation of motion for $C_\mu$ is 
\begin{equation}
	\partial_\mu G^{\mu\nu} = -\kappa e J^\nu + \frac{1}{q}\varepsilon^{ \nu\rho\lambda}\partial_\rho \Sigma_\lambda^{\rm e}\,.
\end{equation}
Let us consider the space components of this equation, i.e. for $\nu = i$\,, where $i=1,2$\,,
\begin{equation}
	\partial_\mu G^{\mu i} = -\kappa e J^i + \frac{1}{q}\varepsilon^{ i \rho\lambda}\partial_\rho \Sigma_\lambda^{\rm e}\,.
\end{equation}
Let us also assume that the sources are independent of time. Then taking the curl of this equation in two dimensions, we find after some algebra,
%
%
%
%
%
%
\begin{equation}
\label{flux}
\boxed{	\epsilon^{ij}\partial_i C_j = - \kappa e \tilde{ J_0}  - \frac{1}{q} \Sigma_0^{\rm e}\,.}
\end{equation}
The second term on the right hand side counts the vortices caused by the external magnetic field. Let us look at the first term more closely by writing $\tilde{ J_0}$ using Gordon decomposition for free fermion fields (and writing $\nabla^2$ for the Laplacian in two space dimensions),
\begin{equation}\label{Gordon}
		\tilde{J^0} = \epsilon^{ij}\frac{1}{-\nabla^2}\partial_iJ_j\\
		 =\epsilon^{ij}\frac{1}{-\nabla^2}\partial_i\left(\frac{i}{2m}\left(\bar{\psi}\partial_j\psi - \partial_j\bar{\psi}\psi\right)+ \frac{1}{m}\partial^k\left(\bar{\psi}\Sigma_{kj}\psi\right)\right)\,.
\end{equation}
We recognize the term inside the parentheses in the first term as the charge current $(J_k)_{\rm cc}$\,. The second term comes from the spin -- $\Sigma_{lk}$ is the spin matrix -- so using the identity  $\Sigma_{lk}=\frac{1}{2}\varepsilon_{lk}\Sigma^3$\,, we can rewrite this equation as
\begin{align}
		\tilde{J^0} 
		& = \frac{1}{2m}\bar{\psi}\Sigma^3\psi + \frac{1}{-\nabla^2}\varepsilon^{ik}\partial_i \left(J_k\right)_{\rm cc}\\
		& \simeq \frac{1}{2m}\psi^\dagger\sigma^3\psi + \frac{1}{-\nabla^2}\varepsilon^{ik}\partial_i \left(J_k\right)^{\rm NR}_{\rm cc}\,,
\end{align}
where in the  last step we have taken the non-relativistic approximation in which the lower component of Dirac spinor is ignored. Thus we may rewrite Eq.~(\ref{flux}) as 
\begin{equation}
	\boxed{\epsilon_{ij}\partial_iC_j = -\frac{\kappa e}{2m}\psi^\dagger\sigma^3\psi + \frac{\kappa e}{\nabla^2}\varepsilon_{ik}\partial_i \left(J_k\right)^{\rm NR}_{\rm cc} - \frac{1}{q}\Sigma_0^{\rm e}\,.}
\end{equation}
Now we proceed to take the { macroscopic} average on both sides. The average of the first term on the right hand side gives the density of magnetic moment $\mu$ in the third direction, so we can write
\begin{equation}
	\langle\epsilon_{ij}\partial_iC_j\rangle = 
	-\kappa \mu  + \frac{\kappa e}{\nabla^2}\varepsilon_{ik}\partial_i\langle\left(J_k\right)^{\rm NR}_{\rm cc}\rangle - \frac{1}{q}\Sigma_0^{\rm e}\,,
\end{equation}
where we have taken $\Sigma_0^{\rm e}$ to represent the density of flux vortices induced by the external magnetic field. 

In the absence of any external electric field, the average current should vanish, $\langle\left(J_k\right)^{NR}_{cc}\rangle=0$\,. Also if the externally applied electric field produces a steady current in a fixed direction, the current should have vanishing curl and hence $\varepsilon_{ik}\partial_i \langle\left(J_k\right)^{NR}_{cc}\rangle =0$\,. 
Thus in general we may write
\begin{equation}\label{flux2}
	\langle\epsilon_{ij}\partial_iC_j\rangle = -{\kappa }\mu	- \frac{1}{q}\Sigma_0^{\rm e}\,.
\end{equation}
The magnetic moment density is associated with the fermions. The external magnetic field can pass through the material only as vortices of quantized flux. Electrons will scatter from these vortices, giving rise to an alignment of electrons along the direction of magnetic field so that $\mu$ becomes non-zero. The average value of magnetic moment per electron $\mu$ depends on the strength of external magnetic fields (hence on vortex density), magnetic susceptibility of the material hosting these electrons and possibly on other properties of the system.

Eq.~(\ref{flux2}) resembles the equation of motion of Chern-Simons gauge theory which expresses that vorticity is locally determined by magnetic moment density $\mu$. Specifically, it suggests that if we approximate the magnetic moment density as $\mu(\vec{x}) = \sum\limits_i \mu_i\delta^2(\vec{x}- \vec{x}_i)$ as a distribution of particles with magnetic moment, the curl of $C_i$ at the $i$-th particle's location is ${\kappa \mu_i}$\,.  As we mentioned earlier, the curl of $C_\mu$ is the magnetic field inside the vortex string. Thus 
we have found the desired phenomenon of flux attachment in this system of vortices and electrons. We shall see in the next section that this leads to fractional statistics of these particles.
%

\section{Fractional Statistics}\label{anyon}
Let us  take the non-relativistic limit of the fermion equation and consider one particle going around another. We start from the reduced Lagrangian containing only the matter field coupled to $C_\mu$\, which we can read off from Eq.~(\ref{Z_B}),
\begin{equation}
	\mathscr{L}_C= \bar{\psi}(i\gamma^\mu \partial_\mu - m)\psi - v^2 C_\mu \frac{1}{\triangle + M^2} \epsilon^{\mu\nu\lambda}\partial_\nu \Sigma_\lambda^{\rm e} - \kappa evM C_\mu \frac{1}{\triangle + M^2} \varepsilon^{\mu\nu\lambda} \partial_\nu \tilde{J_\lambda}- \frac{v^2}{4} G_{\mu\nu}\frac{1}{\triangle + M^2} G^{\mu\nu}\,.
\end{equation}
%
%
%
In the low energy limit we may replace the propagator $\frac{1}{\triangle + M^2}$ by $\frac{1}{M^2}$ to write 
\begin{equation}
	\mathcal{L}= \bar{\psi}(i\gamma^\mu \partial_\mu - m)\psi + \kappa e C_\mu J^\mu- \frac{1}{q} C_\mu  \epsilon^{\mu\nu\lambda}\partial_\nu \Sigma_\lambda - \frac{1}{4} G_{\mu\nu} G^{\mu\nu}\,,
	\label{low-energy-lagrangian}
\end{equation}
where we have used the definition $\tilde{J_\lambda}= \varepsilon_{\lambda\rho\sigma}\partial^\sigma \frac{1}{\triangle}J^\rho $\,, and also  redefined $C_\mu$ as $\frac{1}{q}C_\mu$\,.
We see that in the low energy limit $C_\mu$ is minimally coupled to fermions. The Dirac equation, as derived from the above Lagrangian, is given by
\begin{equation}
	(i\slashed\partial + \kappa e \slashed C - m )\psi = 0\,.
\end{equation}
In order to take the non-relativistic limit of this equation, we first note that the fermion acts as a source of $C_\mu$\,. So for a nearly static fermion the field $C_\mu$ will also be approximately time-independent, and in the approximation that we can treat the electrons as point particles, $\langle\nabla \times \vec{C}\rangle$ is non-zero only at the particle's location. Therefore, another fermion going around the first one held at some point $\vec{x}_0$ sees a time-independent configuration of $C_\mu$ with a non-zero $\langle\nabla \times \vec{C}\rangle$ only at $\vec{x}_0$\,.
Then in the non-relativistic limit $ E - m = E_{NR} \ll m$\,, the Dirac equation reduces to 
\begin{equation}
	\left[-\frac{1}{2m} \left(\vec{\nabla} + i\kappa e\vec{C}\right)^2 + \kappa e \sigma^3 (\vec{\nabla}\times \vec{C})_3 + \kappa e C_0 \right]\psi_A = E_{NR}\psi_A.
\end{equation} 

Let us now take an electron in a closed path $\Gamma$ around a nearly static one. It will pick up a phase 
%
\begin{align}
		\psi(x) &= \psi_0(x) \exp\left(-i \kappa e\oint_{\Gamma} \vec{C}\cdot \vec{dl}\right) \notag\\
		& = \psi_0(x) \exp\left(-i \kappa e \int_{\sigma}\left( \vec{\nabla}\times \vec{C}\right)\cdot \vec{ds}\right) \notag
		\\ & = \psi_0(x)\exp\left(i\kappa^2 e \int_\sigma \mu ds + \frac{i\kappa e}{q}\int_\sigma \Sigma^e_0 ds\right)\,. \label{A-B-effect}
\end{align}
Here we have used Eq.~(\ref{flux2}) and replaced the integral on the right hand side by its average over $\sigma$\,, which is a surface enclosed by $\Gamma$\,.

Since $\Sigma_0^e$ is the flux due to the external field and it is quantized inside the material, let us consider a loop which encloses only one electron but no flux of the external field. Then the second term makes no contribution.
The first term is more interesting. The magnetic moment density $\mu$ vanishes everywhere inside $\Gamma$ except at the location of the electron, so the change in phase of one electron going around another is ${\kappa^2 e}\mu_1\,,$
where $\mu_1$ is the magnetic moment of the electron inside the loop. If both electrons have the same magnetic moment, the phase for exchanging the two electrons is 
\begin{equation}\label{phase-change}
	\Delta\theta = \frac{1}{2}\kappa^2 e \mu_1\,.
\end{equation}
Since $\kappa^2 = 1/L$ is arbitrarily chosen, this exchange phase is not quantized and the electrons with flux attached to them obey fractional statistics, i.e. they become anyons.


\section{Discussion}\label{disc}
%
Usually an investigation of 2+1-dimensional physics starts from a theory defined on a plane. In this paper, we have attempted to describe 2+1-dimensional phenomena by starting in 3+1 dimensions, from the dual of a theory of magnetic flux strings coupled to electrons, and reducing it to one dimension less. The higher dimensional theory shows a linear potential between pairs of electrons, which can be interpreted as the result of a flux string connecting the spin magnetic moments of the two electrons. When reduced dimensionally by freezing all motion in one direction, the strings are reduced to vortices and the magnetic flux at each particle's location is proportional to the magnetic moment of the particle. This indicates a ``flux attachment'' similar to that arising in the effective description of fractional quantum Hall effect through the Chern-Simons term. We also find that similar to the FQHE description, the particles with attached flux exhibit fractional statistics.

In this paper we have also found that spin magnetic moment is the central cause of attachment of flux to particles. This is quite natural because vorticity will appear wherever magnetic field penetrates the superconducting bulk. Thus in our framework such flux attachment or flux pinning would be impossible for spinless (scalar) charged particles, which is in stark contrast with similar phenomena in Chern-Simons theories.  

A brief comparison with Chern-Simons theory may be relevant here. It is well known that the Chern-Simons gauge field has no dynamics of its own when it is not coupled to matter. When a coupling to matter current is added, it constrains the magnetic field to be nonvanishing only at the locations of the charged particles. The electric field in any direction also turns out to be proportional to the transverse current, which is a characteristic feature of Hall effect. We have started from the Abelian Higgs system in the broken phase, with flux strings as well as electrons, and dualized it to a $B\wedge F$-type theory. Unlike pure Chern-simons gauge theory, it is not a purely topological field theory, but when reduced to 2+1 dimensions, the coupling between the vortices and electrons also constrains the magnetic field to be nonvanishing only at the locations of the particles. This phenomenon of flux attachment leads to fractional statistics of the particles just as for Chern-Simons theory, but it will require further investigations to check if Hall effect is also present in the setup that we have considered. 

The attachment of magnetic flux to spin (or more precisely magnetic moment) that we have found above may be compared with the ad hoc construction adopted in~\cite{Balents:PRB61}, where electrons with up and down spins were treated as bosonic operators interacting with a Chern-Simons gauge field which attaches flux to spin. Then in the bosonic variables, a dual theory of vortices could be constructed and different phases could be analyzed. We have shown that flux attachment to spin can occur in a specific type of boson-fermion model, namely the Abelian Higgs model with itinerant electrons; we can expect that similar phases, e.g. chiral spin liquid, or superconductors with tightly bound pairs, will be realized in this system as well.

Finally, we would like to comment on the $\mathscr{L}_3$ which we ignored after Sec.~\ref{dimred}. This is justified in the lowest order, with the fermions being treated as noninteracting, so that $J_3$ is independent of the other two components of the fermion current. In general, interactions will ensure that the effects of $\mathscr{L}_3$ should be taken into account. But there is a special situation when $\mathscr{L}_3$ decouples from $\mathscr{L}_{2+1}$ -- when there is  complete spin polarization of electrons along the external magnetic field. To see this we write $J_3$ by using Gordon's decomposition,
\begin{equation}
	J^3  = \frac{1}{m} \partial_\nu (\bar{\psi}\Sigma^{3\nu} \psi)= \frac{1}{m} \partial_0 (\bar{\psi}\Sigma^{30} \psi) + \frac{1}{m} \partial_i (\bar{\psi}\Sigma^{3i} \psi)\,.
\end{equation}
We assume the fermionic fields to be slowly varying in time. Then the first term is small. The second term is proportional to $\varepsilon^{ij} \partial^i (\psi^\dagger \sigma^j \psi)$, where $\psi$ is the nonrelativistic two-component Pauli spinor. Thus in a fully spin polarized state of fermions, $J^3 \approx 0$ and the two parts decouple from each other. Another important feature of this special state is that in this case the flux attachment equation takes the form 
\begin{equation}
	\langle\epsilon_{ij}\partial_iC_j\rangle = 
	-\frac{\kappa e}{2m} (\rho_\uparrow - \rho_\downarrow ) = -\frac{\kappa e}{2m} \rho\,, 
\end{equation}
where $\rho = \rho_\uparrow$ is the density of particles. Thus we see that in this case the flux attachment equation reduces to the form obtained in CS theory. These observations make the spin polarized system very special and demands further investigation.




\end{document}